\documentclass[pra,twocolumn,showpacs,superscriptaddress,showkeys]{revtex4}

\usepackage{amsmath,amssymb,amsthm,times,graphics,graphicx,bm,hyperref}

\newcommand{\be}{\begin{equation}}
\newcommand{\ee}{\end{equation}}
\newcommand{\ben}{\begin{eqnarray}}
\newcommand{\een}{\end{eqnarray}}
\newcommand{\bes}{\begin{subequations}}
\newcommand{\ees}{\end{subequations}}
\newcommand{\dg}{\dagger}

\newcommand{\norm}[1]{\left\|\,#1\,\right\|}       
\newcommand{\enorm}[1]{\norm{#1}_{\mathrm{2}}}      
\newcommand{\fnorm}[1]{\norm{#1}_{\mathrm {F}}}    

\newcommand{\set}[1]{{\left\{#1\right\}}}    
\newcommand{\pmsmt}{\{\Pi^A_j\}}

\newcommand{\dens}{{D}(\mathcal{H}^M\otimes\mathcal{H}^N)}
\newcommand{\abs}[1]{\left\lvert #1 \right\rvert}

\newcommand{\ro}{\rho}
\newcommand{\rof}{\ro_f}

\newcommand{\dqc}{\rho_{DQC1}}
\newcommand{\discm}{\mathcal{D}(\ro)}

\newcommand{\fu}{\operatorname{d}(\ro,U_A)}
\newcommand{\fua}[2]{\operatorname{d}(#1,#2)}
\newcommand{\fum}{\operatorname{d}_{\operatorname{max}}(\ro)}
\newcommand{\fuma}[1]{\operatorname{d}_{\operatorname{max}}(#1)}
\newcommand{\fumdqc}{\operatorname{d}_{\operatorname{max}}(\dqc)}

\def\ket#1{ | #1 \rangle}
\def\bra#1{{\langle #1 | }}
\newcommand{\ketbra}[2]{\ket{#1}\!\bra{#2}}        
\def\tr{ {\rm{Tr }}}
\newcommand{\braket}[2]{\mbox{$\langle #1  | #2 \rangle$}}
\newcommand{\proj}[1]{\mbox{$|#1\rangle \!\langle #1 |$}}

\newtheorem{theorem}{Theorem}

\begin{document}
\title{Signatures of non-classicality in mixed-state quantum computation}
\author{Animesh Datta}
 \affiliation{Institute for Mathematical Sciences, 53 Prince's Gate, Imperial College, London, SW7 2PG, UK}
 \affiliation{QOLS, The Blackett Laboratory, Imperial College London, Prince Consort Road, SW7 2BW, UK}
\author{Sevag Gharibian}
 \affiliation{Institute for Quantum Computing, University of Waterloo, Waterloo, Canada}

\date{\today}
\begin{abstract}
We investigate signatures of non-classicality in quantum states,
in particular, those involved in the DQC1 model of mixed-state
quantum computation [Phys. Rev. Lett. \textbf{81}, 5672 (1998)].
To do so, we consider two known non-classicality criteria. The
first quantifies disturbance of a quantum state under locally
noneffective unitary operations (LNU), which are local unitaries
acting invariantly on a subsystem. The second quantifies
measurement induced disturbance (MID) in the eigenbasis of the
reduced density matrices. We study the role of both figures of
non-classicality in the exponential speedup of the DQC1 model and
compare them \textit{vis-a-vis} the interpretation provided in
terms of quantum discord. In particular, we prove that a non-zero
quantum discord implies a non-zero shift under LNUs. We also use
the MID measure to study the locking of classical correlations
[Phys. Rev. Lett. \textbf{92}, 067902 (2004)] using two mutually
unbiased bases (MUB). We find the MID measure to exactly
correspond to the number of locked bits of correlation. For three
or more MUBs, it predicts the possibility of superior locking
effects.
\end{abstract}

 \pacs{03.65.Ud,03.67.Mn,03.67.Lx}
 \keywords{Quantum discord, DQC1, Locking}

\maketitle

\section{Introduction}

        A thorough understanding of classical and quantum correlations
underlies their successful exploitation in quantum information
science. The relative roles and abilities of these two forms of
correlations in performing specific computational and information
processing tasks would be a valuable advance in the field.
Substantial progress in this direction have already been achieved.
The role of entangled states in quantum information processing and
computing is quite well studied. Jozsa and Linden \cite{jozsa03a}
showed that multipartite entanglement must grow unboundedly with
the problem size if a pure-state quantum computation is to attain
an exponential speedup over its classical counterpart. In the
context of information processing, Masanes has shown
\cite{masanes06a} that all bipartite entangled states can enhance
the teleporting power of some other state. In spite of these
successes, there are instances of quantum computations where the
quantum advantage cannot be attributed to entanglement. Meyer has
presented a quantum search algorithm that uses no entanglement
\cite{meyer00a}. Instances are also known of oracle based problems
that can be solved without entanglement, yet with certain
advantages over the best known classical algorithms
\cite{biham04a},\cite{kenigsberg06a}.

Given this scenario, it becomes a logical necessity to study the
essentialness of entanglement in quantum information science. The
oldest signature of quantum behavior has been nonlocality.
Interestingly, it is well known that quantum nonlocality and
entanglement are not equivalent notions
\cite{bennett99c},\cite{methot07a}. Entanglement stems from the
superposition principle, or the amplitude description of quantum
mechanics. This description is, however, not one that uniquely
defines quantum mechanics. Consequently, it should not be a
surprise that entanglement cannot capture the whole power of
quantum mechanics. This provides a significant motivation for
studying alternative certificates of quantum behavior.

A much more realistic motivation is that provided by mixed-state
quantum computation. Pure states in a quantum computation
inevitably get mixed due to decoherence. Countering this requires
the techniques of quantum error-correction. A different way to
address this issue would be to study the prospects of quantum
computational speedup with mixed states themselves \cite{asv00}.
NMR quantum computation provides a perfect scenario for this. As a
simplified model for this, Knill and Laflamme proposed the DQC1 or
the `power of one qubit' model \cite{kl98}. Though not believed to
be as powerful as a pure-state quantum computer, it is known to
provide an exponential speedup over the best known classical
algorithm for estimating the normalized trace of a unitary matrix.
The DQC1 model was found to have a limited amount of (bipartite)
entanglement that does not increase with the system size.
Additionally, for certain parameter settings, there is no
distillable entanglement present whatsoever, and yet the model
retains its exponential advantage. In this latter case the state
has a positive partial transpose, and thus possesses, at most,
just bound entanglement \cite{datta05a}. Looking for a more
satisfactory explanation for the exponential speedup, the quantum
discord \cite{ollivier01a},\cite{henderson01a} was calculated, of
which the amount found was a constant fraction of the maximum
possible \cite{datta08a}, regardless of the parameter settings for
the model. In this paper, we study two alternative methods of
studying the quantum behavior of the DQC1 model.

Locally noneffective unitary operations (LNU) have previously been
studied with the aim of developing an entanglement detection
criterion~\cite{f06},\cite{gkb08}. Here, we study the LNU as a
possible notion of non-classicality, motivated by the disturbance
of a quantum state under unitary operations. We provide a brief
introduction to the LNU in Sec~\ref{S:lc}. In Sec~\ref{S:lcdqc1},
we employ LNU in analyzing the DQC1 model. The DQC1 model has
previously been studied using the quantum discord. Thus, in
Sec~\ref{S:discordandlc}, we compare these two certificates of
non-classicality, with the aim of contrasting \emph{disturbance
under measurement} with \emph{disturbance under unitary
operations}. We then move on to study the DQC1 model using the
measurement-induced disturbance (MID) measure~\cite{Luo08a} in
Sec~\ref{S:middqc}. In Ref.~\cite{Luo08a}, a preliminary analysis
of the DQC1 model was begun. Here, we extend this analysis to the
entire parameter range for the DQC1 model, including those which
limit the DQC1 state to being at most bound entangled. This latter
case is of particular interest due to the lack of distillable
entanglement. Later, in Sec~\ref{S:locking}, we present an example
in the realm of quantum communication where the MID measure is a
good certificate of non-classicality. Specifically, we study the
construction of Ref.~\cite{dhlst04} which uses two mutually
unbiased bases (MUB) to lock classical correlations in a quantum
state. The value of the MID measure in this case is exactly the
number of locked bits of correlation in the state. Considering the
same construction with more than two MUBs, the MID measure
portends superior locking abilities, though they must involve MUBs
more general than those based on Latin squares and generalized
Pauli matrices~\cite{BW07}. We conclude with a brief discussion in
Sec~\ref{S:conclusion}.

Throughout, we denote a vector by $\bm{v}$, and take all
logarithms to base $2$. We define $\dens$ as the set of density
operators acting on the $MN$-dimensional Hilbert space
$\mathcal{H}^M\otimes\mathcal{H}^N$. All designations of a density
matrix without any subscripts will be implied to mean a bipartite
state. For example, $\tau$ shall stand for $\tau_{AB}$.

\section{Locally Noneffective Unitary Operations (LNU)}
\label{S:lc}

We begin by introducing locally noneffective unitary operations
(LNU), first proposed under the name local \emph{cyclic}
operations~\cite{f06}. For this, consider a bipartite quantum
state $\ro\in\dens$, shared between $A$ and $B$ such that
$\rho_A=\tr_B(\ro)$ and $\rho_B=\tr_A(\ro)$. Suppose now that
Alice performs a local unitary $U_A$ that does not change her
subsystem, that is, $\rho_A = U_A \rho_A U^{\dag}_A$, or
equivalently
 \be
 \label{lccond}
 [\rho_A,U_A] =0.
 \ee
This action can, however, affect the state of the total system,
such that if we define $\rof:=(U_A \otimes \mathbb{I}_B) \ro(U_A
\otimes \mathbb{I}_B)^{\dag}$, it is possible that $\ro\neq\rof$.
Unitaries satisfying Eqn.~(\ref{lccond}) are called
LNU~\cite{f06}. To quantify the difference between $\ro$ and
$\rof$, we use 
 \ben
  \label{lcmeasure}
 \fum &:=& \max_{\scriptsize\begin{array}{c}
 U_A:\\
  $[$\rho_A$,$U_A$]=0$ \\
\end{array}}
  \frac{1}{\sqrt{2}}\fnorm{\ro-\rof} \nonumber \\
  &=& \max_{\scriptsize\begin{array}{c}
 U_A:\\
  $[$\rho_A$,$U_A$]=0$ \\
\end{array}} \sqrt{\tr(\ro^2)-\tr(\ro\rof)}.
 \een
where $\fnorm{A}=\sqrt{\tr(A^\dg A)}$ denotes the Frobenius norm.
From the latter expression, it is clear that $0\leq \fum \leq 1$.
For any product state $\rho_{prod}:=\rho_A\otimes\rho_B$,
$\fuma{\rho_{prod}}=0$. Closed form expressions for $\fum$ are known for (pseudo)pure states and Werner states~\cite{gkb08}. As with the quantum discord, it is possible to have
$\fuma{\rho_{sep}}>0$ for certain separable states, implying
$\fum$ is not a non-locality measure. A separable state
$\rho_{sep}\in\dens$ is defined as one of the form
\begin{equation}
    \rho_{sep}:=\sum_k p_k\ketbra{a^k}{a^k}\otimes\ketbra{b^k}{b^k},
\end{equation}
where $\sum_k p_k=1$, and the $\ket{a^k}\in\mathcal{H}^M$ and
$\ket{b^k}\in\mathcal{H}^N$ are vectors of Euclidean norm $1$. For
two-qubit separable states, the maximum LNU distance attainable
is~\cite{f06}
\begin{equation}\label{eqn:ccbound}
\fuma{\rho_{sep}}\leq \frac{1}{\sqrt{2}}.
\end{equation}

As an illustration, the maximum LNU distance for the two-qubit
isotropic state,
 \be
 \ro_{iso} = \frac{1-z}{4}I_4 +
 z\proj{\Psi},\;\;\;\;\;z\in [0,1]
  \ee
where $\ket{\Psi}=(\ket{00}+\ket{11})/\sqrt{2}$, 
is given by~\cite{gkb08}
 \be
    \fuma{\ro_{iso}} = z.
 \ee
By Eqn.~(\ref{eqn:ccbound}), we can  conclude that the two-qubit
isotropic state is entangled for $z >1/\sqrt{2}$. The partial
transpose test, which in this case is necessary and sufficient,
shows that this state is actually entangled for all $z>1/3$,
showing that the LNU distance is weaker at detecting
entangled states than the former.

We remark that we have restricted our attention here to the case
where the LNU is applied to subsystem $A$ of $\ro$. Let us derive
a simple upper bound on $\fum$ which holds regardless
of which target subsystem we choose, and which proves useful throughout this paper.

\begin{theorem}\label{thm:fuBound}
    For any $\ro\in\dens$,
    \be
        \fum\leq\sqrt{2\left(\tr(\rho^2)-\frac{1}{MN}\right)}.
    \ee
\end{theorem}
\begin{proof}
   Since $\fnorm{\rho-\frac{I}{MN}}$ is invariant under
unitary operations, we have via the triangle inequality that:
    \ben
        \fnorm{\rho-\rho_f}&\leq& \fnorm{\rho-\frac{I}{MN}} + \fnorm{\frac{I}{MN}-\rho_f} \nonumber \\
        &=&2\fnorm{\rho-\frac{I}{MN}}  \nonumber\\
        &=&2\sqrt{\tr(\rho^2)-\frac{1}{MN}}
    \een
    Substituting this expression in Eqn.~(\ref{lcmeasure}) gives the desired result.
\end{proof}

Thus, if the purity of a state $\ro$ strictly decreases as a
function of the dimension, then $\fum\rightarrow0$ as
$MN\rightarrow\infty$.

\section{LNU in the DQC1 model}
\label{S:lcdqc1}

\begin{figure}
\resizebox{5.5cm}{2.5cm}{\includegraphics{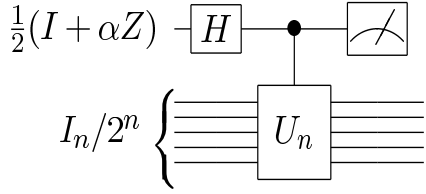}}
 \caption[The `power of one qubit' model]{The DQC1 circuit} \label{F:dqc1}
\end{figure}

        We now study the non-classical features of the DQC1 model
of quantum computation, as quantified by $\fum$. The $n+1$ qubit
DQC1 state, as demonstrated in Fig~(\ref{F:dqc1}), is given
by~\cite{datta05a}
 \be
 \label{E:dqc1}
\dqc =\frac{1}{2^{n+1}}\left(
\begin{array}{cc}
  I_n & \alpha U^{\dg}_n \\
  \alpha U_n & I_n \\
\end{array}%
\right).
 \ee
We will consider the top qubit to be system $A$ on which our local
unitary acts and the remaining $n$ qubits as system $B$. The
reduced state is then
  \be
\rho_A = \tr_B(\dqc) = \frac{1}{2}\left(
\begin{array}{cc}
  1 & \alpha \tau^* \\
  \alpha \tau & 1 \\
\end{array}%
\right)
  \ee
with $\tau = \tr(U_n)/2^n$. 
For an arbitrary $\mathrm{SU}(2)$
unitary $U_A$ acting on $A$
, which we characterize as
\be
 \label{su2}
U_A=\left(
\begin{array}{cc}
  e^{i\phi}\cos\theta & e^{i\chi}\sin\theta \\
  -e^{-i\chi}\sin\theta   & e^{-i\phi}\cos\theta \\
\end{array}%
\right),
 \ee
the LNU condition of
Eqn.~(\ref{lccond}) requires that $\chi =
\frac{\pi}{2}-\arg(\tau)$ and either $\phi =0$ or $\theta=\pi/2$.
Both cases lead to the same final expression, so set
$\phi=0$. Via Eqn.~(\ref{lcmeasure}) and simple algebra, we hence have
 $$
 \fua{\dqc}{\theta} =
 \frac{\alpha\sin\theta}{2^{(n+1)/2}}\sqrt{1-\frac{\mathrm{Re}\tr(e^{-2i\arg\tau}U^2_n)}{2^n}}.
 $$
The now trivial maximization over all $\theta$ gives
 \ben
 \label{dqclc}
 \!\!\!\!\fumdqc\!\! &=&\!\! \frac{\alpha}{2^{(n+1)/2}}\sqrt{1-\frac{\mathrm{Re}\tr(e^{-2i\arg\tau}U^2_n)}{2^n}}\\
            \!\!& \leq & \!\!\frac{\alpha}{2^{n/2}}.
 \label{dqclcfinal}
 \een
Here, we have used the rough estimate
$\mathrm{Re}\tr(e^{2i\arg\tau}U^2_n)\geq -2^n$. For a two-qubit
pure state ($n=1, \alpha=1$), we thus have $\fumdqc \leq
1/\sqrt{2}$, which conforms with Eqn.~(\ref{eqn:ccbound}).
A typical instance
of the DQC1 circuit is provided by that of a random unitary $U_n$
in the DQC1 circuit of Fig~(\ref{F:dqc1}). For such instances of
large enough Haar distributed unitaries, $\tr(U_n^2)$ is bounded
above by a constant with high probability ~\cite{diaconis03}.
Thus, the second term inside the square root in Eqn.~(\ref{dqclc})
is approximately zero, and
 \be
 \fumdqc \approx \frac{\alpha}{2^{(n+1)/2}}.
 \ee

This shows that the DQC1 state experiences very little disturbance
under LNU, and in fact this disturbance vanishes asymptotically as
$n$ grows. As discussed in the introduction, it would appear that
the quantum discord is better suited~\cite{datta08a} to
quantifying non-classicality in the DQC1 model. This, however,
raises the question of how the discord and LNU distance are
related, and whether the paradigms of `disturbance under
measurement' and `disturbance under unitary operations' lead to
differing notions of non-classicality. We explore these questions
in the following section.

Before closing, for completeness, we invoke Theorem~(\ref{thm:fuBound}) to show that the LNU distance is exponentially decreasing for \emph{any} other choice of bi-partitions $A$ and $B$ of the qubits in $\dqc$. In fact, since
 \be\label{eqn:dqcMixedness}
    \tr(\dqc^2)=\frac{1+\alpha^2}{2^{n+1}},
 \ee
Theorem~(\ref{thm:fuBound}) immediately gives the same upper bound
of Eqn.~(\ref{dqclcfinal}). 

\section{Quantum Discord \lowercase{\textit{vs}} LNU Distance}
\label{S:discordandlc}

Motivated by the fact that both the quantum discord and the LNU
distance are aimed at capturing the non-classical features in a
quantum state via an induced disturbance, we seek an answer to the
question of whether one implies the other in any sense or not.
Here, we show that non-zero quantum discord implies a
non-zero LNU distance, but that the converse is not necessarily
true. We begin with a formal definition of quantum discord.

Given a quantum state $\rho\in\dens$, its quantum mutual
information is defined as $\mathcal{I}(\rho) := S(\rho_A) +
S(\rho_B) - S(\rho)$. The quantum mutual information can, however,
also be defined in an inequivalent way as
 \be
\mathcal{J}_{\set{\Pi_j^A}}(\rho) =
S(\rho_B)-S\left(\rho_{B|\set{\Pi_j^A}}\right)
 \ee
with
 $$
    S(\rho_{B|\set{\Pi_j^A}}) = \sum_jp_jS\left((\Pi_j^A\otimes I^B)\rho(\Pi_j^A\otimes I^B)\Big/p_j\right),
 $$
where $p_j=\tr(\Pi_j^A\otimes I^B\rho)$. Projective measurements
on subsystem $A$ removes all non-classical correlations between $A$
and $B$. The quantity $\mathcal{J}$ thus signifies a measure of
classical correlations in the state $\rho$ \cite{henderson01a}. To
ensure that it captures all classical correlations, we need to
maximize $\mathcal{J}$ over the set of one dimensional projective
measurements. This leads to the definition of quantum
discord~\cite{ollivier01a} as
 \ben \label{eqn:discord_def}
    \discm &:=& \mathcal{I}(\rho)-\max_{\set{\Pi_j^A}}\mathcal{J}_{\set{\Pi_j^A}}(\rho) \nonumber \\
           &=&
           S(\rho_A)-S(\rho)+\min_{\set{\Pi_j^A}}S\left(\rho_{B|\set{\Pi_j^A}}\right).
 \een
Intuitively, quantum discord captures purely quantum correlations
in a quantum state. This is distinct from entanglement in the case
of mixed states. For pure states, quantum discord reduces to the
von-Neumann entropy of the reduced density matrix, which is a
measure of entanglement. On the other hand, it is possible for
mixed separable states to have non-zero quantum discord. The main
theorem concerning the discord that we require here is the
following.
\begin{theorem}[Ollivier and Zurek~\cite{ollivier01a}]\label{thm:olzu}
        For $\rho\in\dens$, $\discm=0$ if and only if $\rho = \sum_j
(\Pi_j^A\otimes I^B)\rho(\Pi_j^A\otimes I^B)$, for some complete
set of rank one projectors $\set{\Pi_j^A}$.
\end{theorem}

We now show the following.
\begin{theorem}\label{thm:1}
    For $\rho\in\dens$, if $\discm>0$, then $\fum>0$.
\end{theorem}
\begin{proof}
    We begin by writing $\ro$ in Fano form~\cite{F83}, i.e.
    \ben\label{eqn:fano}
        \ro = &\frac{1}{MN}&(I^A\otimes I^B + \bm{r}^A\cdot\bm{\sigma}^A\otimes{I^B}+\hspace{10mm}\\&&
                    I^A\otimes\bm{r}^B\cdot\bm{\sigma}^B+\sum_{s=1}^{M^2-1}\sum_{t=1}^{N^2-1}T_{st}\sigma^A_s\otimes\sigma^B_t).\nonumber
    \een
Here, $\bm{\sigma}^A$ denotes the $(M^2-1)$-component vector of
traceless orthogonal Hermitian generators of $SU(M)$ (which
generalize the Pauli spin operators), $\bm{r}^A$ is the
$(M^2-1)$-dimensional Bloch vector for subsystem $A$ with
$r^A_s=\frac{M}{2}\tr(\rho_A\sigma^A_s)$, and $T$ is a real matrix
known as the correlation matrix with entries
$T_{st}=\frac{MN}{4}\tr(\sigma^A_s\otimes\sigma^B_t\ro)$. The
definitions for subsystem $B$ are analogous.

An explicit construction for the generators $\sigma_i$ of $SU(d)$
for $d\geq 2$ is given as follows~\cite{he81}. Define
$\set{\sigma_i}_{i=1}^{d^2-1}= \set{U_{pq},V_{pq},W_{r}}$, such
that for $1\leq p<q\leq d$ and $1\leq r \leq d-1$, and
$\set{\ket{k}}_{k=1}^{d}$ some complete orthonormal basis for
$\mathcal{H}^d$:
    \bes
    \begin{eqnarray}
        U_{pq}&=&\ket{p}\bra{q}+\ket{q}\bra{p}\label{eqn:Ugenerators}\\
        V_{pq}&=&-i\ket{p}\bra{q}+i\ket{q}\bra{p}\label{eqn:Vgenerators}\\
        W_{r}\!\! &=&\!\!\!\sqrt{\frac{2}{r(r+1)}}\!\!\left(\sum_{k=1}^{r}\ket{k}\bra{k}-r\ket{r+1}\bra{r+1}\!\!\right)\label{eqn:Wgenerators}
    \end{eqnarray}
    \ees
In our ensuing discussion, without loss of generality, for $SU(M)$
we fix the choice of basis $\set{\ket{k}}_{k=1}^{M}$ above as the
eigenbasis~\footnote{The set of orthonormal eigenvectors of
$\rho_A$ will not be unique if the eigenvalues of $\rho_A$ are
degenerate. Hence, we fix some choice of eigenbasis for $\rho_A$
as the ``canonical'' choice to be referred to throughout the rest
of our discussion.} of $\rho_A$.

Assume now that $\discm>0$. Then, any choice of complete
measurement $\pmsmt$ must disturb $\ro$, i.e. by
Theorem~\ref{thm:olzu}, if we define \be
    \rof:= \sum_{j=1}^M(\Pi_j^A\otimes I)\ro(\Pi_j^A\otimes I),
\ee then $\rof\neq
\ro$~\cite{ollivier01a},\cite{henderson01a},\cite{dattathesis}.
Henceforth, when we discuss the action of $\pmsmt$ on $\rho_A$, we
are referring to the state $\sum_{j=1}^M\Pi_j^A\rho_A\Pi_j^A$.
Now, let $\pmsmt$ be a complete projective measurement onto the
eigenbasis of $\rho_A$. Then, $\pmsmt$ acts invariantly on
$\rho_A$, and thus must alter the last term in Eqn.
(\ref{eqn:fano}) to ensure $\rof\neq\ro$. To see this, recall that
one can write $\rho_A=\frac{1}{M}(I^A +
\bm{r}^A\cdot\bm{\sigma}^A)$, from which it follows that if
$\pmsmt$ acts invariantly on $\rho_A$, then it also acts
invariantly on $\bm{r}^A\cdot\bm{\sigma}^A$ from
Eqn.~(\ref{eqn:fano}). Since all generators
$\sigma^A_s\in\set{W_r}_r$ are diagonal, it follows that there
must exist some $T_{st}\neq 0$ such that
$\sigma^A_i\in\set{U_{pq},V_{pq}}_{pq}$. We now use this fact to
construct a LNU $U^A$ achieving $\fu>0$.

Define unitary $U^A$ as diagonal in the eigenbasis of $\rho_A$,
i.e. $U^A = \sum_{k=1}^{M}e^{i\theta_k}\ket{k}\bra{k}$, with
eigenvalues to be chosen as needed. Then, $[U^A,\rho_A]=0$ by
construction, and so $U^A\otimes I^B$ must alter $T$ through its
action on $\ro$ to ensure $\rof\neq \ro$. Focusing on the last
term from Eqn.~(\ref{eqn:fano}), we thus have:
    \ben\label{eqn:TshiftSetup}
\sum_{s=1}^{M^2-1}\sum^{N^2-1}_{t=1}T_{st}U^A\sigma^A_s{U^A}^\dg\otimes
\sigma^B_t = \hspace{3.2cm}\nonumber \\
\sum_{s=1}^{M^2-1}\sum^{N^2-1}_{t=1}T_{st}
        \Bigg(\sum_{m=1}^{M}\sum_{n=1}^{M}e^{i(\theta_m-\theta_n)}
        \bra{m}\sigma^A_s\ket{n}\ket{m}\bra{n}\Bigg)\otimes
        \sigma^B_t\hspace{-0.7cm}\nonumber
    \een
Analyzing each generator $\sigma^A_s$ case by case, we find, for
some $1\leq p<q\leq M$ or $1\leq r \leq M-1$:
    \ben
        \sum_{m=1}^{M}\sum_{n=1}^{M}e^{i(\theta_m-\theta_n)}\bra{m}\sigma_s\ket{n}\ket{m}\bra{n}=\hspace{3.0cm}\nonumber\\
        \begin{cases}
            \cos(\theta_p -\theta_q)U_{pq}-\sin(\theta_p -\theta_q)V_{pq}\text{\quad if $\sigma_s=U_{pq}$}\label{eqn:TshiftU}\\
            \sin(\theta_p -\theta_q)U_{pq}+\cos(\theta_p -\theta_q)V_{pq}\label{eqn:TshiftV}\text{\quad if $\sigma_s=V_{pq}$}\\
            W_r\hspace{46mm}\text{\quad if $\sigma_s=W_r$}\label{eqn:TshiftW}
        \end{cases}\nonumber
            \een
Denoting by $T^f$ the $T$ matrix for $\rof$, we have:
    \begin{eqnarray}
        T^f_{st}=
            \begin{cases}\label{eqn:Tfinal}
                \cos(\theta_p-\theta_q)T_{st}+\sin(\theta_p-\theta_q)T_{wt}\\
                    \hspace{28mm}\text{if $\sigma_s=U_{pq}$, where $\sigma_{w}=V_{pq}$}\\
                \cos(\theta_p-\theta_q)T_{st}-\sin(\theta_p-\theta_q)T_{wt}\\
                    \hspace{28mm}\text{if $\sigma_s=V_{pq}$, where $\sigma_{w}=U_{pq}$}\\
                T_{st}\hspace{23mm} \text{if $\sigma_s=W_r$}.\\
            \end{cases}\nonumber
    \end{eqnarray}
Thus, if there exists an $s$ such that $T_{st}\neq0$ and
$\sigma^A_s\in\set{U_{pq},V_{pq}}_{pq}$, it follows that one can
easily choose appropriate eigenvalues $e^{i\theta_p}$ and
$e^{i\theta_q}$ for $U^A$ such that $T^f\neq T$, implying
$\fum>0$. By our argument above for $\discm>0$, such an $s$ does
in fact exist.
\end{proof}


To show that the converse of Theorem~\ref{thm:1} does not
hold, we present an example of a zero discord state that has
non-zero LNU measure. Consider the two qubit separable state
 $$
\ro = \frac{1}{2}\left(\frac{{I}_2 +
\bm{a}.\bm{\sigma}}{2}\otimes\frac{{I}_2 + \bm{b}.\bm{\sigma}}{2}
+ \frac{{I}_2 - \bm{a}.\bm{\sigma}}{2}\otimes\frac{{I}_2 -
\bm{b}.\bm{\sigma}}{2} \right),
 $$
where $\enorm{\bm{a}}=\enorm{\bm{b}}=1$. This state, by
construction, has zero discord for a single qubit measurement on
either $A$ or $B$. To see this, consider the projective
measurements
    $$
    \set{\frac{{I}_2 \pm
\bm{a}.\bm{\sigma}}{2}}
   $$
on $A$. Let us now study the LNU distance for this state, with the
local unitary being applied to say $A$. Notice that $\rho_A =
\rho_ B = {I}_2/2$, and $\tr(\ro^2)=1/2$. The former implies that
the set of allowed local unitaries is the whole of $SU(2)$, an
element of which is given by Eq (\ref{su2}). Let us for
convenience parameterize $\bm{a} = (0,0,1)$ and $\bm{b} =
(\sin\gamma \cos \delta, \sin\gamma\sin \delta, \cos\gamma)$.
Then, some algebra leads to
 \be
 \tr(\ro\rof)=\frac{1}{2}\cos^2\theta.
 \ee
whose minimum is 0, whereby
 \be
 \fum=\frac{1}{\sqrt{2}}.
 \ee

We thus have an example of a class of separable, zero discord
states which demonstrates a non-zero shift under LNU. In fact, it
attains the maximum shift possible for two-qubit separable states.
Hence, if one wishes to define notions of non-classicality in
quantum states in terms of `disturbance under measurement' versus
`disturbance under unitary operations', and one chooses discord
and the LNU distance as canonical quantifiers of such effects,
respectively, then the resulting respective notions of
non-classicality are not equivalent. As we have shown in
Thm.~\ref{thm:1}, however, the quantum discord is a stronger
notion of non-classicality than the LNU criterion. 

\section{Measuring correlations via Measurement-Induced Disturbance}
\label{S:middqc}

        The measure we intend to use in this section was presented
by Luo in~\cite{Luo08a}. It relies on the disturbance of a quantum
system under a generic measurement. In that sense, it is similar
in spirit to quantum discord, but not quite. In the case of
quantum discord, as per Eqn.~(\ref{eqn:discord_def}), one
maximizes over one-dimensional projective measurements on one of
the subsystems. For the new measure, which we will call the
measurement-induced disturbance (MID) measure, one performs
measurements on \emph{both} the subsystems, with the measurements
being given by projectors onto the eigenvectors of the reduced
subsystems. Then the MID measure of quantum correlations for a
quantum state $\rho\in\dens$ is given by~\cite{Luo08a}
 \be
 \mathcal{M}(\ro) := \mathcal{I}(\ro) - \mathcal{I}(\mathcal{P}(\ro))
 \ee
where
 \be
 \label{E:midstate}
 \mathcal{P}(\ro):=\sum_{i=1}^M\sum_{j=1}^N(\Pi_i^A\otimes\Pi_j^B)
\ro(\Pi_i^A\otimes\Pi_j^B).
 \ee
Here $\{\Pi_i^A\},\{\Pi_j^B\}$ denote rank one projections onto
the eigenbases of $\rho_A$ and $\rho_B$, respectively.
$\mathcal{I}(\sigma)$ is the quantum mutual information, which is
considered to the measure of total, classical and quantum,
correlations in the quantum state $\sigma$. Since no optimizations
are involved in this measure, it is much easier to calculate in
practice than the quantum discord or the LNU distance, which
involve optimizations over projective measurements and local
unitaries respectively. The measurement induced by the spectral
resolution leaves the entropy of the reduced states invariant and
is, in a certain sense, the least disturbing. Actually, this
choice of measurement even leaves the reduced states
invariant~\cite{Luo08a}.
Interestingly, for pure states, both the quantum discord and the
MID measure reduce to the von-Neumann entropy of the reduced
density matrix, which is a measure of bipartite entanglement.

\begin{figure}
 \resizebox{9.5cm}{6cm}{\includegraphics{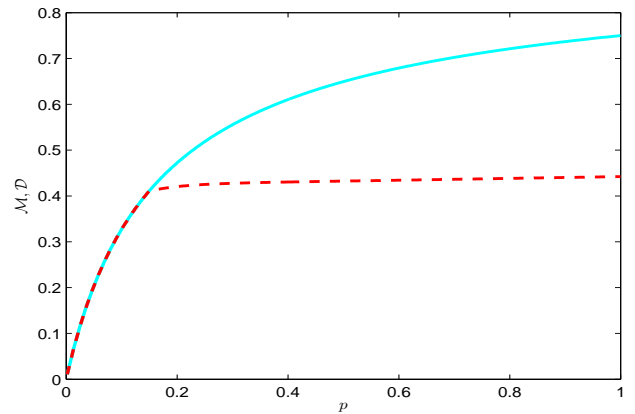}}
\caption{(Color online) The solid line is the MID measure
$\mathcal{M}$ for the $2\times 4$ Horodecki state from~\cite{h97}.
The dashed line is the quantum discord $\mathcal{D}$ for the same
state~\cite{dattathesis}. The kink in the latter curve occurs at
$p=1/7.$ We see here, as in the case of the DQC1 state, that the
MID measure is greater than or equal to the quantum discord.}
\label{Horod}
\end{figure}
                As a nontrivial example, we will consider the well-known Horodecki
bound entangled state in $2\otimes 4$ dimensions~\cite{h97}. It is
bound entangled for all values of $0\leq p \leq 1$, and the state
is given as
 \be
\rho_H = \frac{1}{1+7p}\left(\!\!\!\!%
\begin{array}{cccccccc}
  \;p\;&\;0\; &\;0\;& \;0\; & \;0\; & \;p\; & \;0\; & \;0\; \\
  0 & p & 0  & 0 & 0 & 0 & p& 0 \\
  0 & 0 & p  & 0 & 0 & 0 & 0 & p \\
  0 & 0 & 0  & p & 0 & 0 & 0 & 0 \\
  0 & 0 & 0  & 0 & \frac{1+p}{2} & 0 & 0 & \frac{\sqrt{1-p^2}}{2} \\
  p & 0 & 0  & 0 & 0 & p & 0 & 0 \\
  0 & p & 0  & 0 & 0 & 0 & p & 0 \\
  0 & 0 & p  & 0 &\frac{\sqrt{1-p^2}}{2} & 0 & 0 & \frac{1+p}{2}  \\
\end{array}%
\!\!\!\!\right).\nonumber
 \ee
From this, the projectors onto eigenvectors of the reduced density
matrices can be calculated to be
 \ben
\{\Pi^A_1,\Pi^A_2\}&=&\left\{ \left(%
\begin{array}{cc}
  1 & 0 \\
  0 & 0 \\
\end{array}%
\right),
\left(%
\begin{array}{cc}
  0 & 0 \\
  0 & 1 \\
\end{array}%
\right)
\right\},\;\;\;\;\;\;\mbox{and}\nonumber\\
\{\Pi^B_1,\cdots,\Pi^B_4\} &=&
\big\{\proj{\Psi^{+}},\proj{\Psi^{-}},\nonumber\\
        &&\hskip1.0cm\proj{\Phi^{+}},\proj{\Phi^{-}}
\big\}. \nonumber
 \een
where $\ket{\Psi^{\pm}}=(\ket{1}\pm \ket{2})/\sqrt 2$ and
$\ket{\Phi^{\pm}}=(\ket{0}\pm \ket{3})/\sqrt 2,$ with
$\{\ket{0},\ket{1},\ket{2},\ket{3}\}$ forming the computational
basis for the second subsystem. Using these in Eqn.
(\ref{E:midstate}), we can easily obtain
  $$
\mathcal{P}(\rho_H)= \frac{1}{1+7p}\!\!\left(\!\!\!\!%
\begin{array}{cccccccc}
  \;p\;&\;0\; &\;0\;& \;0\; & \;0\; & \;0\; & \;0\; & \;0\; \\
  0 & p & 0  & 0 & 0 & 0 & 0& 0 \\
  0 & 0 & p  & 0 & 0 & 0 & 0 & 0 \\
  0 & 0 & 0  & p & 0 & 0 & 0 & 0 \\
  0 & 0 & 0  & 0 & \frac{1+p}{2} & 0 & 0 & \frac{\sqrt{1-p^2}}{2} \\
  0 & 0 & 0  & 0 & 0 & p & 0 & 0 \\
  0 & 0 & 0  & 0 & 0 & 0 & p & 0 \\
  0 & 0 & 0  & 0 &\frac{\sqrt{1-p^2}}{2} & 0 & 0 & \frac{1+p}{2}  \\
\end{array}%
\!\!\!\!\right).
 $$
This density matrix is different from the original one in that
there are no coherences between the two subsystems. The MID
measure for this state can then easily be obtained analytically as
$\mathcal{M}(\rho_H)=S(\mathcal{P}(\rho_H))-S(\rho_H)$ and is
plotted in Fig (\ref{Horod}). In the same figure is shown the
quantum discord for this state, when a measurement is made on the
two-dimensional subsystem. For the details of its calculation, see
Ref.~\cite{dattathesis}. As we see, there are non-classical
correlations in this state that are not distillable into maximally
entangled Bell pairs. Another instance, dealt with next, is the
DQC1 state, which for $\alpha < 1/2$ is, at best, bound entangled,
having failed to show any entanglement by partial transposition
criterion across any bipartite split. It even failed to show any
entanglement at the second level of the scheme
of~\cite{doherty04a}. It therefore might be possible to the
quantify the intrinsic information processing abilities of these
bound entangled states using the measures dealt with in this
paper.

\subsection{MID measure in the DQC1 model}

\begin{figure}
 \resizebox{9.5cm}{6cm}{\includegraphics{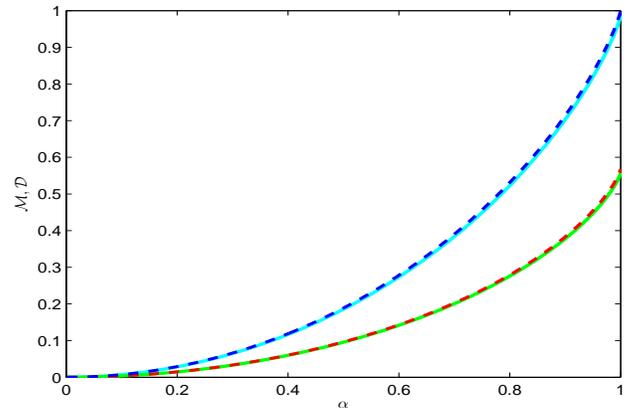}}
\caption{(Color online) The upper solid (cyan) line is the MID
measure $\mathcal{M}$ (Eqn.~(\ref{mid})) for the DQC1 circuit for
a $n=5$ qubit Haar distributed random unitary matrix. The upper
dashed (blue) line is the analytic expression for the MID measure
for DQC1 states with a  Haar distributed random unitary matrix
(Eqn. (\ref{E:midanal})). The lower dashed (red) line shows the
discord $\mathcal{D}$ in the DQC1 circuit with the same unitary.
The lower solid (green) line shows the analytical expression in of
quantum discord from~\cite{datta08a}. All quantities are shown as
functions of the purity of the control qubit. } \label{measdisc}
\end{figure}

We now move on to calculate the MID measure in the DQC1 model. Our
analysis extends that of~\cite{Luo08a}, where only the case of
$\alpha=1$ was considered. Considering $\alpha<1/2$ here will be
of particular interest, due to the lack of distillable
entanglement in the DQC1 state. Consequently, we start with the
$n+1$ qubit DQC1 state, given by Eqn.~(\ref{E:dqc1}), wherefrom
 \be
 \rho_{A}= \frac{1}{2}\left(
\begin{array}{cc}
  1 & \alpha \tau^* \\
  \alpha \tau & 1 \\
\end{array}%
\right)\;\;\;\;\;\mbox{and}\;\;\;\; \rho_{B}= I_n/2^n.
   \ee
  The projectors onto their respective eigenvectors are
   $$
\{\Pi_1^A,\Pi_2^A\}= \left\{\frac{1}{2}\left(
\begin{array}{cc}
  1 & e^{-i\phi} \\
  e^{i\phi} & 1 \\
\end{array}%
\right),\frac{1}{2}\left(
\begin{array}{cc}
  1 & -e^{-i\phi} \\
  -e^{i\phi} & 1 \\
\end{array}%
\right)\right\}
   $$
where $\tau=re^{i\phi}$ for $r=\abs{\tau}$ is the normalized trace
of $U_n$, i.e. $\tau = \tr(U_n)/2^n$, and
  $$
 \{\Pi_j^B\}=\{E_j\}\;\;\;\;\;
 \mbox{where}\;\;\;[E_j]_{kl}=\delta_{kj}\delta_{lj},\;\;j,k,l=1,\cdots,2^n.
  $$
Using this, we can calculate
 \ben
 \mathcal{P}(\dqc)&=&\sum_{j=1}^{2^n}\sum_{i=1}^2(\Pi_i^A\otimes\Pi_j^B)\dqc(\Pi_i^A\otimes\Pi_j^B) \nonumber\\
                       &=&\frac{1}{2^{n+1}}\sum_j\left(
                                                        \begin{array}{cc}
                                                        1 & \alpha d_j \\
                                                      \alpha d^*_j & 1 \\
                                                        \end{array}%
                                                            \right)\otimes E_j \nonumber\\
                        &=& \frac{1}{2^{n+1}}\left(
                                                    \begin{array}{cc}
                                                    I_n & \alpha D \\
                                                   \alpha D^{\dg} & I_n \\
                                                   \end{array}%
                                                   \right)
 \een
where $d_j=(u_{jj}^*+e^{-2i\phi}u_{jj} )/2$, with $u_{jj}$ being
the $(j,j)$th entry of $U_n$, and
 $$
D = \mathrm{diag}\left(d_{1},\cdots,d_{j},\cdots\right).
 $$ Since $D$ is diagonal,
it is fairly easy to obtain the spectrum of $\mathcal{P}(\dqc)$,
which is given by
 \be
{\bm \lambda}[\mathcal{P}(\dqc)]=\left\{\frac{1\pm \alpha
|d_i|}{2^{n+1}}\right\} \;\;\;\;\mbox{for}\;\;\;i=1,\cdots,2^n.
 \ee
Letting $\lambda_k$ denote the $k$th entry of ${\bm
\lambda}[\mathcal{P}(\dqc)]$, the von-Neumann entropy of this
state is
 \ben
 S(\mathcal{P}(\dqc)) &=&-\sum_{k=1}^{2^{n+1}} \lambda_k\log(\lambda_k)\nonumber\\
    &=& n+1 -\frac{1}{2^{n+1}}\sum_{j=1}^{2^n}\Bigg(\log(1-\alpha^2|d_j|^2) \nonumber\\
    &+&\alpha|d_j|\log\left(\frac{1+\alpha|d_j|}{1-\alpha|d_j|}\right)\Bigg).
 \een
Now,
  \be
 S(\dqc)= n+H_2\left(\frac{1-\alpha}{2}\right),
  \ee
and the entropies of the partial density matrices being identical,
 \ben
 \label{mid}
\mathcal{M}_{DQC1}\!\! &=&\!\! \mathcal{I}(\dqc)-\mathcal{I}(\mathcal{P}(\dqc))\nonumber \\
            &=&\!\! S(\mathcal{P}(\dqc))-S(\dqc) \nonumber\\
            &=&\!\! 1\!-\! H_2\!\!\left(\frac{1-\alpha}{2}\right)\!\! -\! \frac{1}{2^{n+1}}\!\sum_i\!\Bigg(\!\!\log(1-\alpha^2|d_i|^2) \nonumber\\
            && +\;\;\alpha|d_i|\log\left(\frac{1+\alpha|d_i|}{1-\alpha|d_i|}\right)\Bigg).
 \een
Here, $\abs{d_i}=\abs{u_{ii}\cos(\phi+\beta_i)}$ where
$u_{ii}=re^{i\beta_i}$ for $r=\abs{u_{ii}}$. Given a unitary,
which is known in any implementation of the DQC1 circuit, the
above quantity can be computed easily. Not surprisingly, if the
random unitary is diagonal, the measure $\mathcal{M}$ for the DQC1
circuit actually reduces to its quantum discord (seen via Eqns.
(12) and (13) of~\cite{datta08a}). For a Haar distributed random
unitary matrix, $|u_{ii}| \sim 1/2^{n/2}.$ In the asymptotic limit
of large $n$, $|d_i|\rightarrow 0$, in which case the whole
quantity within the summation in Eqn. (\ref{mid}) goes to zero.
Then,
 \be
 \label{E:midanal}
\mathcal{M}_{DQC1} = 1-H_2\left(\frac{1-\alpha}{2}\right).
 \ee
One fact immediately notable is that the above expression for the
MID measure is independent of $n$, for large $n$. The result for a
$n=5$ qubit Haar distributed random unitary matrix is shown in Fig
(\ref{measdisc}). As is evident, despite the approximations used
in the derivation of Eqn. (\ref{E:midanal}) the asymptotic
analytic expression matches the numerical result at $n=5$ quite
well.

    The MID measure for the DQC1 state across the bipartite split
separating the top qubit from the rest is non-zero for all
non-zero values of the polarization. Across this split, the DQC1
state is strictly separable~\cite{datta05a} and possesses no
entanglement. Hence, it is natural to propose the MID measure as a
quantifier of the resource behind the quantum advantage in the
DQC1 model~\cite{Luo08a}. As can be seen from Fig.
(\ref{measdisc}), the behavior of the MID measure is qualitatively
quite similar to that of the quantum discord. To argue that one is
behind the quantum advantage in the DQC1 model as opposed to the
other would be quite premature. Though both these measures attempt
to capture the quantum feature of disturbance under measurement,
they are quantitatively quite different. We will come back to this
point in Section~\ref{S:conclusion}.

\subsection{MID measure in quantum communication}
\label{S:locking}

We now present an example where the MID measure can be used to
interpret the locking of classical correlations in quantum states.
It has been shown~\cite{dhlst04} that there exist bipartite
quantum states which contain a large amount of locked classical
correlation which can be unlocked by a small amount of classical
communication. More precisely, there exist $2n + 1$-qubit states
for which the optimal classical mutual information between
measurement results on the subsystems can be increased from $n/2$
bits to $n$ bits via a single bit of classical communication.
Despite the impossibility of this feat classically, the states
used in the protocol are not entangled.

Here we use the MID measure to explain this purely quantum
phenomenon. To do so, we evaluate the former on a generalization
of the state used in~\cite{dhlst04},
 \be
 \rho=\frac{1}{md}\sum_{k=1}^{d}\sum_{t=1}^{m}(\proj{k}\otimes\proj{t})_A\otimes(\proj{b_k^t})_B,
 \ee
where the set of $m$ orthonormal bases
$\set{\set{\ket{b_k^t}}_{k=1}^d}_{t=1}^{m}$ is mutually unbiased
(MUB), i.e. $\forall _{t\neq
t^{\prime},i,j}\braket{b_i^t}{b_j^{t^\prime}}=1/\sqrt{d}$. As in
Ref.~\cite{dhlst04}, when $d=2^n$ and $m=2$, the initial
correlations in this state amount to $n/2$ bits, and by Alice's
sending one bit (the bit $t$) to Bob, they end up with $n+1$
correlated bits. The state being separable, it has no
entanglement. Consequently, we cannot ascribe to it the advantage
exhibited by this protocol.

To calculate the MID measure of this state, we need the reduced
states given by
$$
\rho_A = \frac{I_{md}}{md},\;\;\;\;\rho_B =
\frac{I_d}{d}.
$$
  The eigenvectors are trivially obtained, and $\mathcal{P}(\rho)$
is simply the diagonal of $\rho.$ Thus,
 \be
{\bm
\lambda}[\mathcal{P}(\rho)]=\frac{1}{md}\big\{\!\underbrace{1,\cdots,1}_{d},\underbrace{1/d,\cdots,1/d}_{(m-1)d^2},\underbrace{0,0,\cdots,0}_{d(d-1)}\big\}\nonumber
 \ee
whereby
 \be
S(\mathcal{P}(\rho)) = \log m +(2-\frac{1}{m})\log d.
 \ee
The spectrum of $\rho$ is given by
$$
{\bm\lambda}[\rho]=\frac{1}{md}\big\{\!\underbrace{1,1,\cdots,1}_{md},\underbrace{0,0,\cdots,0}_{md(d-1)}\big\}
$$
which leads to
 \be
S(\rho) = \log m+\log d.
 \ee
Finally, we have
 \be
\mathcal{M}(\rho) = S(\mathcal{P}(\rho)) - S(\rho) =
\left(1-\frac{1}{m}\right)\log d, \label{eqn:lockMID}
 \ee
which for $d=2^n$ and $m=2$ is the exactly equal to the gain
attained by this scheme. Moreover, once Bob receives Alice's bit,
the MID measure for their post-communication state drops to $0$,
the latter being diagonal in a local product basis. This suggests
that the MID measure quantifies exactly those non-classical (yet
not entanglement-based) correlations in $\rho$ which were
initially locked.

A few remarks are in order. Eqn.~(\ref{eqn:lockMID}) suggests that
a better locking effect is possible for $m>2$. However, explicit
constructions to date using more than two MUBs have been unable to
achieve superior locking~\cite{BW07}, suggesting that the choice
of construction for the MUBs plays an important role. In contrast,
Eqn.~(\ref{eqn:lockMID}) holds irrespective of the specific choice
of MUBs. It is also known that if the bases above are constructed
using a large set of random unitaries chosen according to the Haar
measure, then the classical mutual information in $\rho$ between
Alice and Bob can indeed be brought down to a
constant~\cite{HLSW04}. There is also numerical evidence (Appendix
of Ref.~\cite{dhlst04arxiv}) that the dimension of the systems may
play a role in achieving better locking. Further connections
between locking and the MID measure are being investigated.

Finally, for completeness, we remark that
$\tr(\rho^2)=1/(2^{n+1})$, and so by Theorem~\ref{thm:fuBound},
the LNU distance for $\rho$ is bounded by

\be
    \fum \leq \frac{\sqrt{2^n-1}}{2^n}\approx\frac{1}{2^{n/2}}.
\ee Thus, in contrast to the MID measure, the LNU distance once
again reveals vanishing non-classicality with growing $n$.

\section{Conclusions}
\label{S:conclusion}

In this paper, we have analyzed two possible quantifiers of
non-classical correlations beyond quantum entanglement,
specifically locally noneffective unitary operations~\cite{f06},
and the measurement-induced disturbance measure~\cite{Luo08a}, and
compared them to the quantum discord~\cite{ollivier01a} within the
context of the DQC1 circuit~\cite{kl98}.

        The LNU distance showed (Eqn.~(\ref{dqclcfinal})) that
there is little non-classicality in the $n+1$ qubit DQC1 state.
This behavior is very similar to that of negativity in the DQC1
model which was used to characterize its
entanglement~\cite{datta05a}. The crucial difference is that the
bipartite split chosen in Sec~\ref{S:lcdqc1} is separable, and
therefore exhibits no entanglement at all. As the LNU distance
vanishes exponentially quickly with growing $n$, one is
hard-pressed to relegate the role of the resource exponentially
speeding up the DQC1 model to it. Similarly, the LNU distance
suggests vanishing non-classicality in the case of locking of
classical correlations in quantum states. This does not, however,
prove that this kind of quantum characteristic cannot be the
resource behind other forms of quantum advantage.

        The MID measure, on the other hand, is considerably more
satisfactory. The zero-entanglement split in the DQC1 model is
shown to have a non-zero amount of non-classicality as per the MID
measure. The magnitude of this measure, as shown in
Fig.~(\ref{measdisc}), is a constant fraction of its maximum
possible value. The maximum possible value, which is independent
of the size of the system under consideration, is
$\mathcal{M}_{max}=1$, and is attained for the maximally entangled
state. Indeed, for a perfectly pure top qubit $\alpha=1$, the DQC1
state attains this value. The MID measure can thus be ascribed to
be a quantifier of the correlations behind the speedup of the DQC1
model. Indeed, this has already been proposed in~\cite{Luo08a}. Further, the MID measure also performs well in quantifying non-classicality in the scenario of locking classical correlations in quantum states.
The measure, however, lacks a clear physical interpretation of
the form of quantum discord, which motivates its operational
significance as a measure of pure quantum
correlations~\cite{zurek03b}. Further studies in this direction
are required before a comprehensive conclusion can be reached.

\section*{Acknowledgements}

AD thanks Carl Caves and Anil Shaji for numerous stimulating
discussions. AD was supported in part by the US Office of Naval
Research (Grant No. N00014-07-1-0304) and also by EPSRC (Grant No.
EP/C546237/1), EPSRC QIP-IRC and the EU Integrated Project (QAP).
SG was partially supported by Canada's NSERC, CIAR and MITACS. We
also thank the anonymous referee for raising certain points that
led to improvements in the paper.

\bibliography{lcbib}

\end{document}